\documentclass[epj]{svjour}
\usepackage{graphics}
\begin{document}
\title{In-medium properties of ${\bf \it D}$-mesons at FAIR}
\author{L. Tol\'os\inst{1,2}, J. Schaffner-Bielich \inst{3} \and H. St\"ocker\inst{1,3}}                
\institute{FIAS, J.W. Goethe-Universit\"at, Max-von-Laue 1, 60438 Frankfurt (M), Germany \and  Gesellschaft f\"ur Schwerionenforschung, Planckstrasse 1, 64291 Darmstadt, Germany \and Institut f\"ur Theoretische Physik, J.W. Goethe-Universit\"at, Max-von-Laue 1, 60438 Frankfurt (M), Germany}
\date{Received: date / Revised version: date}
%
\abstract{
We obtain the $D$-meson spectral density at finite temperature for the conditions of density and temperature expected at FAIR. We perform a self-consistent coupled-channel calculation taking, as a bare interaction, a separable potential model.  The $\Lambda_c$ (2593) resonance is generated dynamically. We observe that the $D$-meson spectral density develops a sizeable width while the quasiparticle peak stays close to the free position. The consequences for the $D$-meson production at FAIR are discussed.
\PACS{{}
      {14.40.Lb, 14.20.Gk, 21.65+f}   
     } 
} 
\maketitle
\section{Introduction}
The properties of hadrons with charm content under extreme conditions of density and temperature are matter of recent interest in connection with the future CBM experiment (Compressed Baryonic Matter) at the FAIR project (Facility for Antiproton and Ion Research)  at GSI \cite{fair}. In particular, the medium modifications of $D$-mesons can have important consequences for $J/\Psi$ suppression \cite{NA501} as well as  open-charm enhancement in nucleus-nucleus collisions \cite{NA50e}. 

The $J/\Psi$ absorption in an hadronic environment due to the inelastic comover scattering can be modified by the in-medium properties of $D$-mesons. Furthermore, the predicted open-charm enhancement in A+A collisions due to an attractive mass shift for $D$-mesons in the nuclear medium \cite{cassing} may explain the dimuon enhancement observed by the NA50 Collaboration in Pb+Pb collisions \cite{NA50e}. However, the latest results on dimuon production by NA60 \cite{NA60} seem to disregard this possibility. Finally, the $D$-mesic nuclei, which are predicted by the quark-meson coupling (QMC) model \cite{qmc}, result from an attractive $D$-meson potential.

Theoretical predictions based on the QMC model \cite{qmc}, QCD sum-rule (QSR) \cite{arata} and chiral models \cite{amruta} obtain attractive mass shifts  of -50 MeV to -200 MeV at zero temperature for nuclear matter saturation density $\rho_0$, where $\rho_0=0.17 \, {\rm fm}^{-3}$, although a second analysis using QSR predicts only  a splitting of $D^+$ and $D^-$ masses of 60 MeV at $\rho_0$ \cite{weise}. However, the $D$-meson spectral density in dense matter is not studied in those approaches. In our previous work \cite{Tolos04}, the $D$-meson spectral density is obtained by including coupled-channel effects as well as the dresssing of the intermediate propagators. The attractive potential felt by the $D$-meson is found to be strongly reduced or become slightly repulsive \cite{Tolos04}. This statement has been recently supported by new coupled-channel calculations \cite{Lutz05,Angels} based on an improved bare interaction in the charm sector, which has been saturated by a t-channel vector-meson exchange.

In this paper, the $D$-meson spectral density is obtained for the conditions of density and temperature expected for the CBM experiment at FAIR \cite{fair} by performing a self-consistent coupled-channel calculation at finite temperature. The medium effects at finite temperature like the Pauli blocking on the nucleons or the Bose distribution for pions, and the dressing of the $D$-mesons, nucleons and pions are investigated. We conclude that the broadening of the $D$-meson is the only source of open-charm enhancement at FAIR \cite{Tolos06}, contrary to previous expectations based on a strong $D$-meson mass shift \cite{cassing}.

\section{Formalism}

The in-medium $D$-meson self-energy and, hence, the $D$-meson spectral density at finite temperature are obtained by performing a self-consistent coupled-channel calculation taking, as bare interaction, a separable potential for the $s$-wave $DN$ interaction. The parameters of this potential, such as the coupling constant and cutoff, are determined by fixing the position and the width of the $\Lambda_c(2593)$ resonance (see \cite{Tolos04}). Finite temperature effects modify the Pauli blocking of the nucleons or the Bose distribution on the pionic intermediate states as well as the dressing of the $D$-mesons, nucleons and pions. Then, the in-medium $DN$ interaction  at finite temperature reads
\begin{eqnarray}
&&\langle M_1 B_1 \mid G(\Omega,T) \mid M_2 B_2 \rangle = \langle M_1 B_1
\mid V \mid M_2 B_2 \rangle  \nonumber \\
&&+\sum_{M_3 B_3} \langle M_1 B_1 \mid V \mid
M_3 B_3 \rangle \frac {F_{M_3 B_3}(T)}{\Omega-E_{M_3}(T) -E_{B_3}(T)+i\eta} \nonumber \\ 
&&\hspace{1cm}\langle M_3
B_3 \mid
G(\Omega,T)
\mid M_2 B_2 \rangle \ ,
   \label{eq:gmat1}
\end{eqnarray}
where $V$ is the separable potential and $\Omega$  the  starting energy. Here $M_i$ and $B_i$  represent the possible mesons ($D$,$\pi$,$\eta$) and baryons ($N$,$\Lambda_c$,$\Sigma_c$), respectively. The function $F_{M_3 B_3}(T)$ for the $D N$ states corresponds to the Pauli operator, i.e  $Q_{D N}(T)=1-n(k_N,T)$, where
$n(k_N,T)$ is the nucleon Fermi distribution at the corresponding
temperature. The function $F_{M_3 B_3}(T)$ is
$1+n(k_{\pi},T)$ for  $\pi \Lambda_c$ or $\pi \Sigma_c$ states , with $n(k_{\pi},T)$ being the Bose distribution of pions at a given temperature, while it is unity for the other intermediate states.

The in-medium properties of the intermediate states are also modified at finite temperature. For nucleons, we use a temperature-dependent Walecka-type $\sigma-\omega$ model with density-dependent scalar and vector coupling constants \cite{Tolos02}. The self-energy of pions in nuclear matter at finite temperature results from  adding to a small repulsive and constant $s$-wave part, the $p$-wave contributions coming from  the coupling to 1$p$-1$h$, 1$\Delta$-1$h$ and 2$p$-2$h$ excitations modified by short-range correlations (see Appendix of \cite{Tolos02}).

The $D$-meson  potential at a given temperature is then given by
\begin{eqnarray}
&&U_{D}(k_{D},E_{D},T)= \int d^3k_N \ n(k_N,T) \times \nonumber \\ &&  \langle D N \mid
 G_{D N\rightarrow D N} (\Omega = E_N+E_{D},T) \mid D N \rangle \ .
\label{eq:self}
\end{eqnarray}
This is a self-consistent problem for the $D$-meson potential, since the in-medium $DN$ interaction depends on the $D$-meson single-particle energy, which in turn depends on the $D$-meson potential.
Once self-consistency is achieved for the on-shell value
$U_{D}(k_{D},E_D,T)$, the self-energy is obtained according to 
\begin{eqnarray}
\Pi_D(k_D,\omega,T)=2\sqrt{m_D^2+k_D^2} \, U_{D}(k_D,\omega,T) \ ,
\end{eqnarray}
and the corresponding spectral density is
\begin{eqnarray}
S_{D}(k_{D},\omega,T) = - \frac {1}{\pi} \mathrm {Im\,} \frac {1}{\omega^2-m_{D}^2-k_{D}^2
-\Pi_D(k_D,\omega,T)} \ 
\label{eq:spec}
\end{eqnarray}

\section{In-medium $D$-mesons at FAIR}

\begin{figure}
\resizebox{0.5\textwidth}{!}{\includegraphics{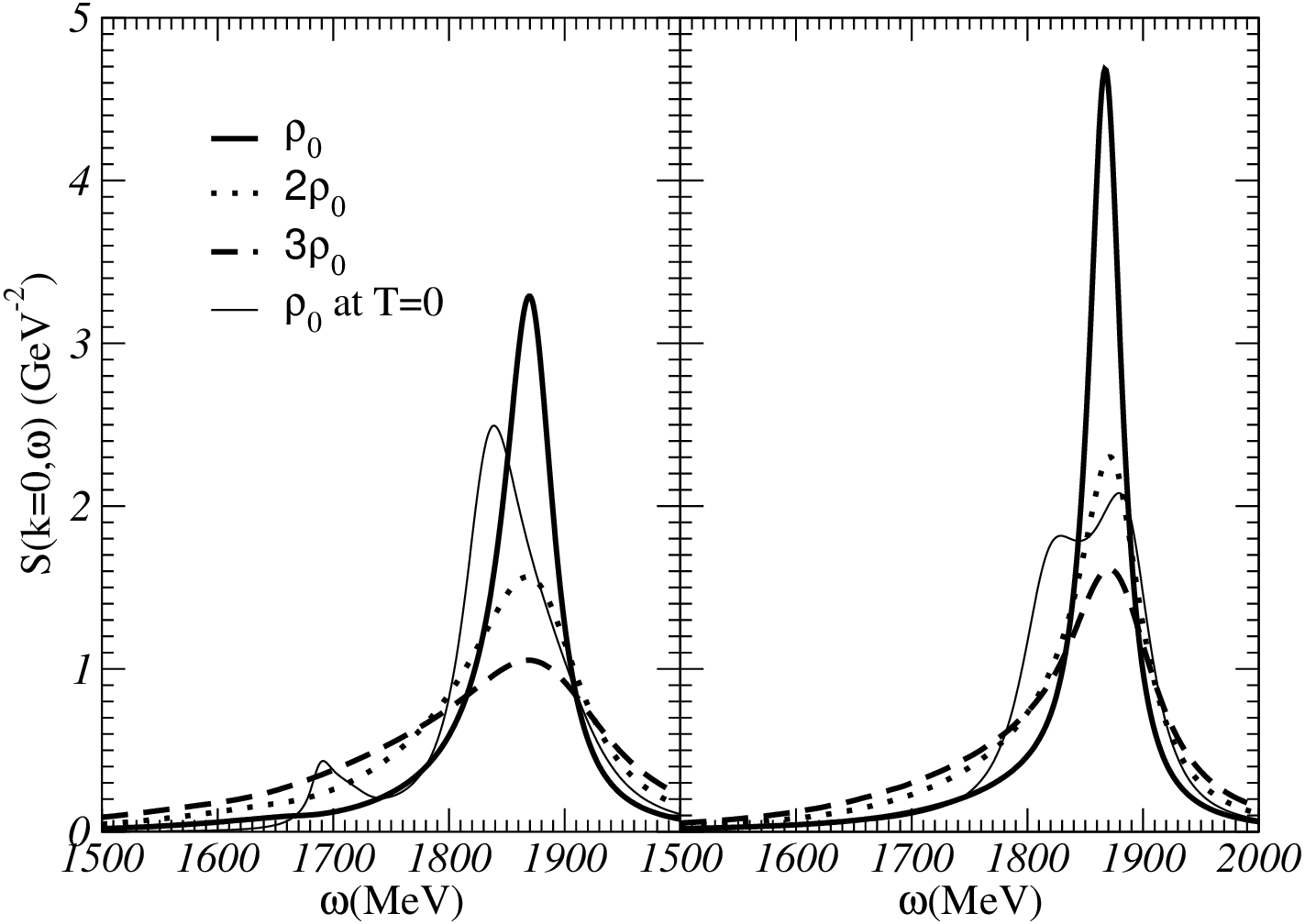}}
\caption{The $D$-meson spectral density at $k_D=0$ and $T=120$ MeV as a function of the energy for different densities and the two approaches discussed in the text. The $D$-meson spectral density at  $k_D=0$ and $T=0$ MeV for $\rho_0$ is also displayed.}
\label{fig:dmeson1}
\end{figure}

In this section we study the  $D$-meson spectral density for the expected conditions of density and temperature in the CBM experiment at FAIR.

 Figure \ref{fig:dmeson1} displays the $D$-meson spectral density at zero momentum and $T=120$ MeV for different densities and for $\Lambda=1$ GeV and $g^2=13.4$. This set of parameters reproduces the position and width of the $\Lambda_c(2593)$ resonance, as seen in \cite{Tolos04}. The temperature is in agreement with the expected temperatures for $D$-meson production at FAIR. Two approaches are considered: self-consistent calculation of the $D$-meson self-energy including the dressing of the nucleons in the intermediate states (left panel) and the self-consistent calculation also including the self-energy of pions (right panel). For comparison we also display the spectral density at $T=0$ for nuclear matter saturation density $\rho_0$.

At finite temperature the quasiparticle peak of the spectral density for $k_D=0$, which is defined as $E_{qp}^2=m_D^2+2\,m_D\,{\rm Re}U_D(k_D=0,E_{qp},T)$,  stays closer to the free position for the range of densities analyzed, compared to the $T=0$ calculation. This is due to the smearing of the Pauli blocking effects with temperature. Temperature also dilutes the structure present at $T=0$ in the spectral distribution at $s \sim 1700$ MeV for the first approach and close to the quasiparticle peak for the second one. Both structures have the $\Lambda_c$-like quantum numbers and correspond to the different behaviour of the in-medium $\Lambda_c(2593)$ resonance  for the two different approaches, as reported in \cite{Tolos04}. While the in-medium $\Lambda_c(2593)$ lies close to its free position for the first approach, the dressing of pions gives rise to two resonant structures in the $I=0$ sector, one of them clearly distinguishable since it lies close to the quasiparticle peak.  However, these structures are washed out with increasing temperature and the $D$-meson spectral density shows then a sizeable width.

Previous works at zero temperature based on the QMC model \cite{qmc}, QSR rules \cite{arata} or chiral effective lagrangians \cite{amruta} predict a strongly attractive $D$-nucleus potential. Our results for $T=0$, however, show that the coupled-channel effects result in an important reduction of the in-medium modifications and  are responsible for the significant width of the $D$-meson spectral density. This effect is independent of the in-medium properties of the intermediate states, as seen in Figure 1. Recent studies of the $D$-meson spectral distribution at $T=0$ suggest that the presence of a new resonance, the $\Sigma_c(2770?)$, results in an even repulsive in-medium $D$-meson potential \cite{Lutz05,Angels}. Finite temperatures effects  make the quasiparticle peak get closer to the $D$-meson free mass and $D$-mesons shows then a considerable width.

\begin{figure}
\resizebox{0.5\textwidth}{!}{\includegraphics{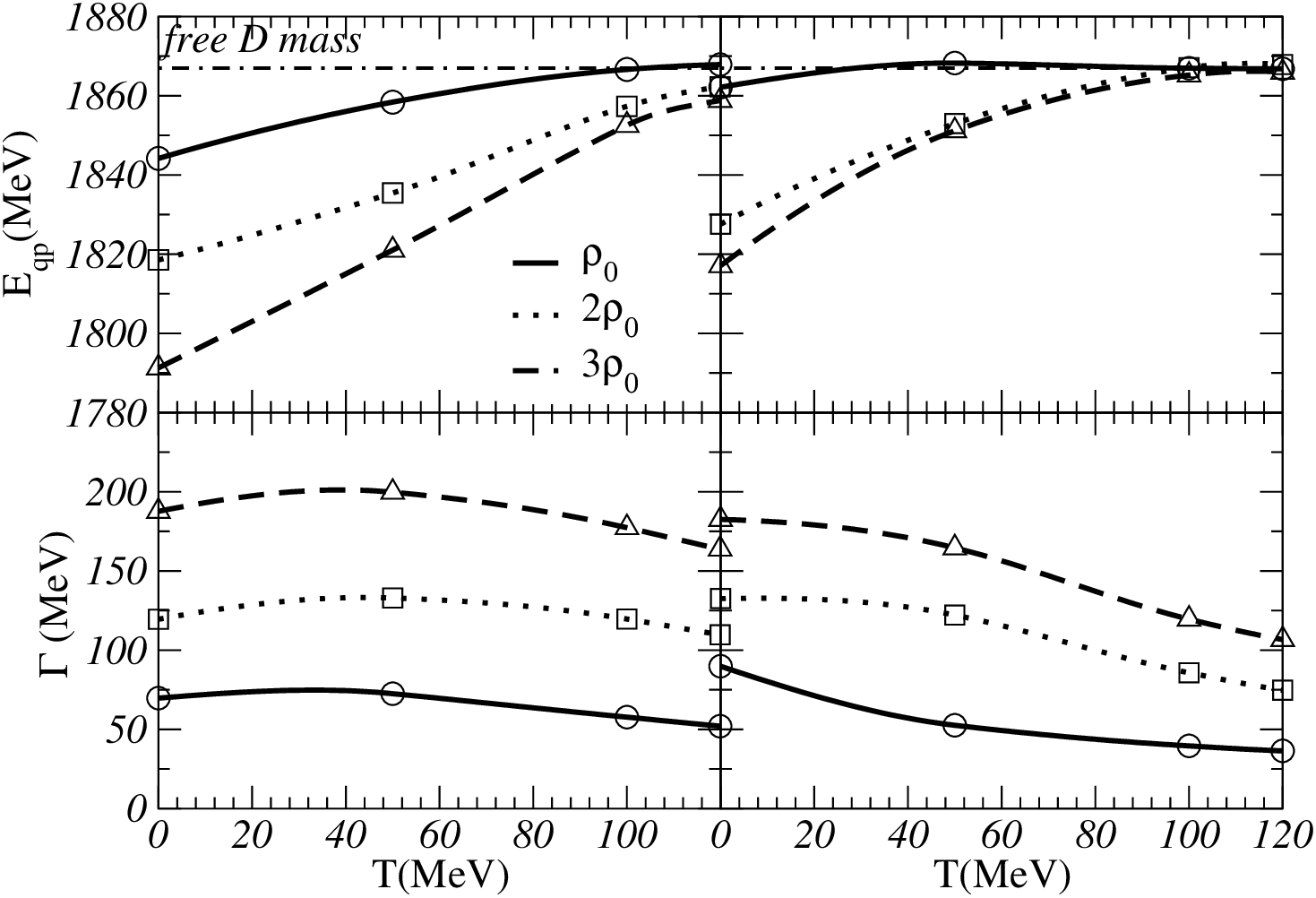}}
\caption{Quasiparticle energy and width of the $D$-meson spectral density at $k_D=0$ as a function of the temperature for different densities and the two approaches discussed in the text.}
\label{fig:dmeson2}
\end{figure}

Figure \ref{fig:dmeson2} shows the quasiparticle energy together with the width of the $D$-meson spectral density at zero momentum, defined as $\Gamma=-{\rm Im}\Pi_D(k_D=0,E_{qp},T)/(2 m_D)$, as a function of the temperature for the previous densities and for the two approaches considered before. For $T=0$ we observe an attractive potential of -23 MeV for $\rho_0$  and -76 MeV for $3\rho_0$ when $D$-mesons and nucleons are dressed in the intermediate states (upper left panel). For the full self-consistent calculation (upper right panel), the $D$-meson potential at $T=0$ lies between -5 MeV for $\rho_0$ and -48 MeV for $ 3 \rho_0$. For higher temperatures, the quasiparticle peak gets close to the $D$-meson free mass, so there is almost no mass shift at finite temperature. On the other hand, the width of the spectral density depends weakly on the temperature and increases strongly with density. At  $T=120$ MeV the width increases from 52 MeV to 163 MeV for $\rho_0$ to $3\rho_0$ for the first approach (lower left panel) and from 36 MeV at $\rho_0$ to 107 MeV at $3 \rho_0$ for the second approach (lower right panel).

The reduced attraction of the $D$-meson in a hot and dense nuclear environment together with a large width can have important consequences for the $D$-meson production in the CBM experiment at FAIR. An enhancement of open-charm in A$+$A collisions was predicted for SPS energies in order to understand the enhancement of 'intermediate-mass dileptons' in Pb$+$Pb collisions \cite{cassing}. A similar effect could be then expected for the conditions of density and temperature at FAIR. However, this prediction was based on the strong mass shift obtained from a QCD sum-rule calculation \cite{arata}. According to our model, the inclusion of a considerable width of the $D$-meson in the medium at finite temperature (40-50 $\rho/\rho_0$ for $T=120$ MeV) would be the only source of enhanced in-medium $D$-meson production, as also seen for antikaons \cite{Tolosratio}. 

\begin{figure}
\resizebox{0.5\textwidth}{!}{\includegraphics{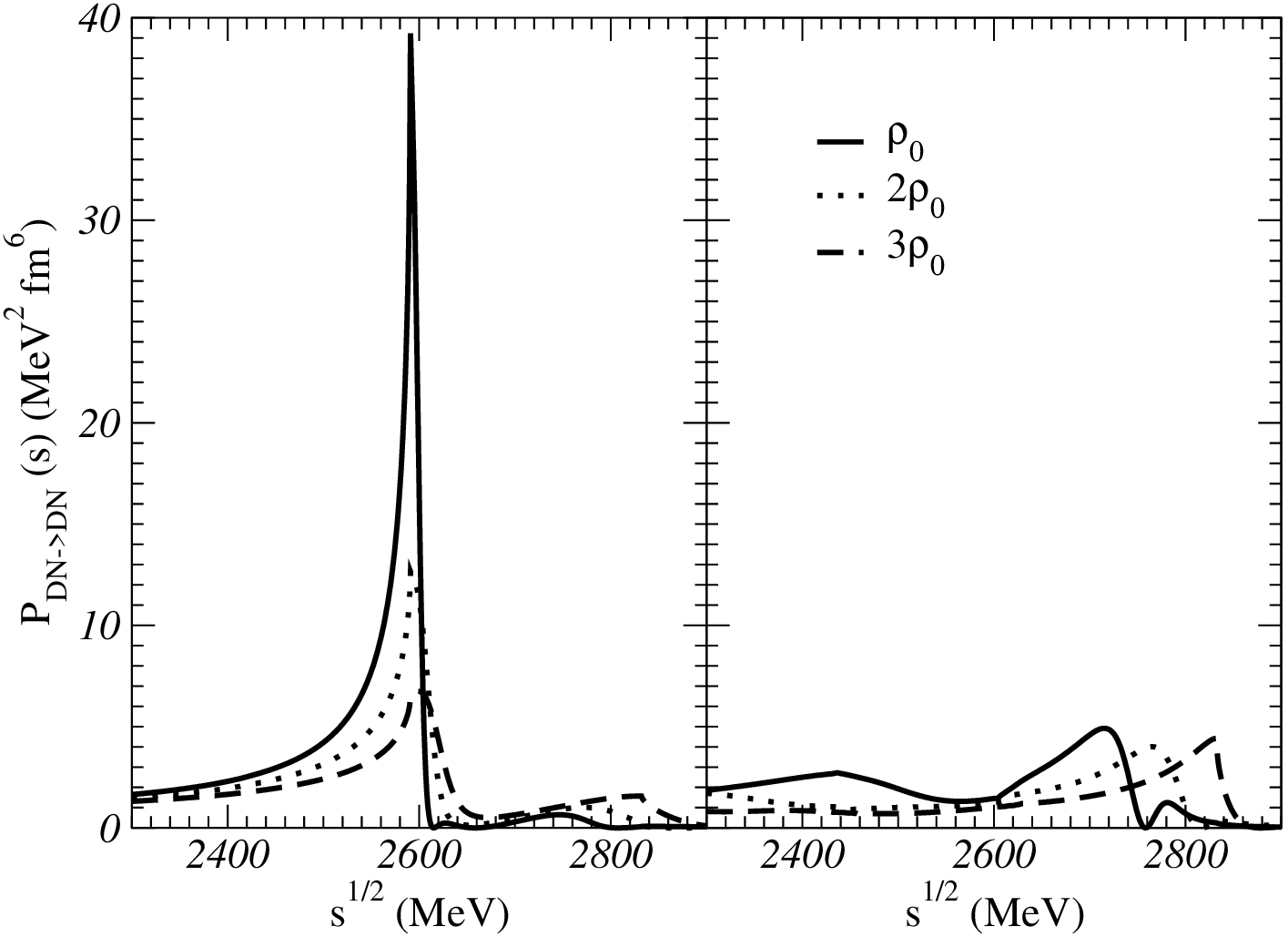}}
\caption{In-medium transition rates for $D^+ n$ ($D^0 p$) at $T=120$ MeV as a function of the center-of-mass energy for different densities and the two approaches discussed in the text.}
\label{fig:dmeson3}
\end{figure}

Another fundamental ingredient for the $D$-meson production in A$+$A collisions is the determination of the $D$-meson cross sections close to the $DN$ threshold. In Figure \ref{fig:dmeson3} we display the elastic in-medium transition rates for $D^+ n$ ($D^0 p$) from $\rho_0$ to $3\rho_0$  at $T=120$ MeV as a function of the center-of-mass energy with and without the inclusion of the pion self-energy. The in-medium transition rates are calculated from the in-medium $DN$ interaction at finite temperature according to
\begin{eqnarray}
\label{crossp} &&P_{D_i+N_i \rightarrow D_f+N_f}(s) = \nonumber \\  
&&\int d\cos(\theta) \
\frac{1}{(2s_{D_i}+1)(2s_{N_i}+1)} \sum_i \sum_\alpha \ G^\dagger G \ ,
\end{eqnarray}
where $D$ and $N$ represent the initial and final $D$-meson and nucleon states. The sums over $i$ and
$\alpha$ run over initial and final spins, while
$s_{D_i}, s_{N_i}$ are the spins of the particles in the entrance channel. 
Once the transition rates are known, the cross sections can be easily determined multiplying by the phase space available. 

For the self-consistent calculation where the $D$-mesons and nucleons are dressed (left panel), we find an enhanced transition rate for energies around the $\Lambda_c(2593)$-resonance mass. The maximum enhancement is reduced by a factor four as density increases from $\rho_0$ to $3 \rho_0$ because of the dilution of the $\Lambda_c(2593)$ with density. The inclusion of the pion dressing in the self-consistent process (right panel) reduces the transition rates drastically compared to the previous approach. We observe the two resonant states with $\Lambda_c$-like quantum numbers for energies around 2.45 GeV and 2.7 GeV. The cross sections close to the $DN$ threshold are expected on the order of 1 to 20 mb in the range of densities studied for both approaches.

\section{Conclusions}

We have studied the in-medium properties of $D$-mesons for the conditions of density and temperature expected in the future CBM experiment at FAIR (GSI). The $D$-meson spectral density in hot and dense nuclear matter is obtained by performing a microscopic self-consistent coupled-channel calculation assuming a separable potential for the $s$-wave $DN$ interaction. The parameters of this potential model have been fitted to generate dynamically the $\Lambda_c(2593)$ resonance.

The $D$-meson spectral density at finite temperature develops a considerable width while the quasiparticle peak stays close to its free position. This small shift of the $D$-meson mass in the nuclear medium at finite temperature is in stark contrast with the large changes (-50 to -200 MeV) claimed in previous mean-field works at zero temperature \cite{qmc,arata,weise,amruta} and is in line with recent zero-temperature self-consistent coupled-channel calculations \cite{Lutz05,Angels}. The self-consistent coupled-channel effects result in a reduction of the in-medium mass shift and a considerable width, independently of the in-medium properties of the intermediate states. 

Mesons with charm content at beam energy close to threshold will be investigated by the CBM experiment \cite{fair}.
Our results imply that the effective mass of the $D$-meson, however, may not be drastically modified in dense matter at finite temperature, but $D$-mesons develop an important width in this hot and dense environment. Therefore, the $D$-meson production in nucleus-nucleus collisions should be obtained in off-shell transport models where also in-medium $D$-meson cross-sections play a crucial role.

\section*{Acknowledgments}

LT. acknowledges financial support from AvH Foundation. This work was supported in part by the Virtual Institute VH-VI-041 of the Helmholtz Association.

\end{document}